\documentclass[aps,prl,preprint,superscriptaddress]{revtex4}
\usepackage{graphicx}
\usepackage{verbatim}

\begin{document}

\title{Observation of magnetization dynamics, Coulomb perturbation and interaction-induced phases on a topological insulator surface}

\author{L. A. Wray}
\affiliation{Department of Physics, Joseph Henry Laboratories, Princeton University, Princeton, NJ 08544, USA}
\affiliation{Princeton Center for Complex Materials, Princeton University, Princeton, NJ 08544, USA}
\affiliation{Advanced Light Source, Lawrence Berkeley National Laboratory, Berkeley, California 94305, USA}
\author{Y. Xia}
\author{S.-Y. Xu}
\affiliation{Department of Physics, Joseph Henry Laboratories, Princeton University, Princeton, NJ 08544, USA}
\author{R. Shankar}
\affiliation{Department of Physics, Joseph Henry Laboratories, Princeton University, Princeton, NJ 08544, USA}
\author{Y.S. Hor}
\author{R.J. Cava}
\affiliation{Department of Chemistry, Princeton University, Princeton, NJ 08544, USA}
\author{A. Bansil}
\author{H. Lin}
\affiliation{Department of Physics, Northeastern University, Boston, MA 02115, USA}
\author{M.Z. Hasan}
\affiliation{Department of Physics, Joseph Henry Laboratories, Princeton University, Princeton, NJ 08544, USA}\affiliation{Princeton Center for Complex Materials, Princeton University, Princeton, NJ 08544, USA}\affiliation{Princeton Institute for Science and Technology of Materials, Princeton University, Princeton, New Jersey 08544, USA}\affiliation{Advanced Light Source, Lawrence Berkeley National Laboratory, Berkeley, California 94305, USA}



\date{Submitted to Nature in December 2009}

\maketitle

\textbf{Three dimensional topological insulators embody a newly discovered state of matter characterized by conducting spin-momentum locked surface states that span the bulk band gap \cite{Intro,TIbasic,DavidNat1,DavidScience,Matthew2008,DavidTunable,MatthewNatPhys}. This highly unusual surface environment provides a rich ground for the discovery of novel physical phenomena and device applications \cite{DavidScience,dhlee,Matthew2008,WrayCuBiSe,Mndoping, BiTeSbTe,FerroPosMassBiSe,zhangTImagImp,FuNew,FerroSplitting,
FuHexagonal,TopoFieldTheory,Biswas,ExcitCapacitor,palee,
ZhangDyon,KaneDevice,DavidNat1,MatthewNatPhys,ZhangPred,
DavidTunable,YazdaniBack}. So far, most of the study on topological insulator surfaces has been limited to understanding their properties without strong Coulomb perturbation or breaking of time reversal symmetry. Here we present the first systematic study of the topological insulator surfaces under strong Coulomb, magnetic and disorder perturbations. Just as the quantum nature of the universe is studied on energy scales of primordial particle formation, the Dirac surface states of topological insulators are defined on the fundamental energy scales of the insulating band gap and bulk electronic dispersions of the underlying spin-orbit matrix. Understanding the response of a topological surface to perturbations on that scale represents a new frontier in uncovering the critical and emergent behaviors of topological surfaces. We have used deposited iron, with a large positive ionization state and significant magnetic moment as a strong probe to modify the surface electronic structure of the Bi$_2$Se$_3$ surface at the gap energy scale. We observe that such perturbation leads to the creation of multiple Dirac fermions consistent with Z$_2$ or Mod(2) symmetry, and that magnetic interactions break time reversal symmetry, allowing the reduction of ungapped surface bands to an even number of species. We present a first-principle theoretical model to account for the observed properties of the altered topological Z$_2$ surface. Taken collectively, these results are a critical guide in manipulating topological surfaces for probing fundamental physics or developing device applications.}

Bismuth selenide has been experimentally discovered by angle resolved photoemission spectroscopy (ARPES) to be a topological insulator with a large bulk band gap ($\sim$300 meV) \cite{MatthewNatPhys,DavidTunable}. Spin resolved photoemission studies reveal that surface electrons in Bi$_2$Se$_3$ form a Dirac cone spanning the bulk insulating gap, composed of spin-momentum locked helical states (Fig. 1a). The Fermi level of grown crystals is usually found to be located in the bulk conduction band due to selenium vacancy defects, however it was subsequently shown that with Ca doping or NO$_2$ surface deposition, the Fermi level can be placed at the Dirac point reaching the topological transport regime \cite{DavidTunable} and magnetic interactions can be controlled via magnetic ions such as Fe \cite{Matthew2008,DavidTunable} or Mn \cite{Mndoping, BiTeSbTe} as shown by Xia \textit{et.al.}, Hsieh  \textit{et.al.}, and Hor  \textit{et.al.}, \cite{Matthew2008,DavidTunable,Mndoping, BiTeSbTe}. Placing the Fermi level at the Dirac point in the presence of magnetic impurities in the bulk can lead to a small gap at the Fermi level (Fig. 1b), however the full character of this gap cannot be decisively resolved due to the lineshape broadening effects. So far, no systematic surface deposition of magnetic impurities to elicit a large systematic magnetic response and bring about controlled changes in the surface band structure has been explored. Probing the effect of magnetic perturbation on the surface is more relevant for potential applications than the previous studies of bulk dopants, because topological insulators need to be in contact with large moment ferromagnets and superconductors for device applications. The focus of this Letter is the exploration of topological insulator surface electron dynamics in the presence of magnetic, charge and disorder perturbations from deposited iron on the surface.

It is known that the surface band structure of Bi$_2$Se$_3$ consists of a single spin-polarized Dirac cone spanning the bulk band gap \cite{MatthewNatPhys}, which can be easily doped with carriers by dissolving small amounts of copper or calcium into the crystal during growth \cite{WrayCuBiSe,HorCa,DavidTunable}. Another route to tuning the chemical potential without greatly changing surface band structure is through surface doping with a sub-monolayer coating of weakly ionized non-magnetic molecules such as NO$_2$ \cite{DavidTunable,MatthewTuneBiTe} (see Fig. 1b). However, topological insulators integrate into devices upon interfacing with strongly perturbing three dimensional materials such as ferromagnets, superconductors in the presence of strongly charged dopants \cite{FerroSplitting,KaneDevice}. Therefore, it is important to probe the microscopic response of topological Z$_2$ or Mod(2) invariant surfaces under strong perturbations, which can potentially significantly modify the surface band structure. In order to achieve such a condition, we deposit iron on the cleaved (111) selenium surface plane and study the systematic changes in the surface band structure. Topological insulator Bi$_2$Se$_3$ cleaves on the selenium surface. The chemistry of iron deposition on Se surfaces is known. Iron deposited on a Se$^{2-}$ surface forms a mild chemical bond, occupying a large ionization state between 2$^+$ and 3$^+$ with roughly 4 $\mu_B$ magnetic moment \cite{IronValence}. As seen in Fig. 1d the surface electronic structure after heavy iron deposition is greatly altered in some specific ways. Five (odd number of) surface bands intersect the Fermi level rather than just one, and these surface states extend below the Fermi level to higher binding energies in the form of multiple Dirac cones.

We studied the surface band structure as a function of deposition level, in order to spectroscopically observe the processes by which iron deposition changes the surface band topology. Fig. 2 shows this evolution as a function of deposition time in crystals with slightly varied chemical compositions of (Sample $\#$1) Bi$_{1.9975}$Ca$_{0.0025}$Se$_3$, (Sample $\#$2) Bi$_2$Se$_{3.04}$, and (Sample $\#$3) Bi$_2$Se$_3$, chosen to contrast the effect of surface perturbations applied when there are different degrees of occupancy in the bulk electronic states. The rate of iron deposition was similar to 4$\%$ of a monolayer per minute. The electronic state prior to deposition is presented in the left-most panel in each doping series, and varies from a slight p-type bulk character with significant bulk resistivity in Sample $\#$1 ($\rho$ = 23 m$\Omega$cm) to n-type bulk metallicity from Bi vacancy defects in as-grown Bi$_2$Se$_3$ (Sample $\#$3). In each case, it is observed that the presence of positively charged Fe surface ions progressively lowers the energy of the surface state and causes the appearance of new surface states with energy/momentum contours similar to the bottom of the bulk conduction band.

The changes brought on by iron deposition can be seen most strikingly in measurements on the bulk p-type Sample $\#$1. Because the bulk chemical potential is near the top of the valence band, the first Dirac cone to lower into view as iron is added can be traced to be the same Dirac cone that can be seen in as-grown Bi$_2$Se$_3$ (labeled D0). After approximately six minutes of deposition, new surface states that do not exist in unperturbed samples become visible at the Fermi level, with new Dirac points labeled D1 and D2. It is also after approximately six minutes of deposition that a gap begins to be apparent at the D0 Dirac node, evidenced by a parabolic shape near the Dirac node, separating the upper and lower Dirac cones of the original surface state. This gap can be seen clearly at a range of incident photon energies in Fig. 1e, confirming that it is a feature of the surface and not the bulk electronic band structure. Electron velocities (band slope) near the D1 and D2 Dirac points increase monotonically as iron is added, showing that iron is increasing the ``Rashba" interaction term (($\vec{k}\times\hat{z})\cdot\vec{\sigma}$, with $\vec{\sigma}$ representing the Pauli matrices) identified in theoretical models \cite{TIbasic}. The number of surface bands intersecting the Fermi level between the $\overline{\Gamma}$- and $\overline{M}$-points progresses from one to three to five, with one band contributed by the original (D0) Dirac cone and two more bands contributed by each of the new (D1,D2) Dirac points. This is consistent with the Mod(2) character of topological surfaces that a topological surface likes
to maintain odd number of Dirac states. After 12 minutes of deposition, the binding energy of the D0 Dirac point was found to have sunk approximately 0.6eV in energy, and the electron binding energies ceased to change under additional deposition. Contrary to the usual trend of deposition experiments, in which the photoemission images become increasingly blurry as molecules are haphazardly added to the surface, we identify a regime in which the image becomes qualitatively sharper with the increasing coverage of iron (Fig. 2a, panels 3-5), corresponding approximately to the increasing clarity of the D0 gap. This observation is important in understanding the mechanism of time reversal symmetry breaking on the surface (discussed with respect to Fig. 4d).

When the chemical potential is positioned near or above the bulk conduction band minimum, as in Samples $\#$2-3 (Fig. 2b,c), all of the bulk band gap is visible within the photoemission image, and we can directly observe the appearance of the new surface state under a low level of deposition. A new surface state appears centered at the Brillouin zone center in Fig. 2b after only 2.5 minutes of deposition, but follows a parabolic (massive) dispersion and is therefore likely to be composed of two weakly split, non-degenerate bands \cite{FuHexagonal}. The new surface band that appears in Sample $\#$3 is clearly strongly split (Fig. 2c(right)), with a Dirac-point at the $\overline{\Gamma}$-point. The fact that an identical deposition time results in strong spin splitting of the D1 band for Sample $\#$3 but not for Sample $\#$2 may suggest a significant aspect of the underlying mechanism by which band splitting comes about in topological surface states. It is known that surface states become spin polarized due to the breaking of parity symmetry (spatial inversion) at the crystal surface, and Rashba spin splitting in topologically trivial systems is likewise associated with a strong potential gradient approaching the cleaved crystal face. Enhanced splitting in the D1 surface state of bulk-metallic Sample $\#$3 likely comes from the steep potential gradient caused by bulk screening of the surface charge, which is expected to occur over just a few nanometers in metallic as-grown Bi$_2$Se$_3$ samples \cite{WrayCuBiSe}. Some of the discrepancy could also be due to a degree of uncertainty in the deposition flow rate calibration from sample to sample, which may have $\sim$20$\%$ variability.

A generalized gradient approximation (GGA) simulation of non-magnetic surface Coulomb perturbation on Bi$_2$Se$_3$ is shown in Fig. 3a, and qualitatively reproduces the progressive appearance of new Dirac points with increasing iron deposition. Through comparison with our numerical result, we can see that the experimentally observed surface states begin to pair off at momentum separation greater than $\sim$0.1$\AA^{-1}$ from the Brillouin zone center, with the upper D0 Dirac cone approaching degeneracy with the lower D1 band, and the upper D1 band connecting to the lower D2 band. At larger momenta, the bands are nearly degenerate with their spin-opposed partner (see Fig. 3b), and follow a dispersion almost identical to the bottom of the conduction band. In Fig. 3c, we have used traced dispersions from the three different samples examined in Fig. 2 to piece together an approximate map of the experimental surface evolution as the surface Coulomb perturbation is varied over a large range. The deposition-induced change in the binding energy of the D0 Dirac point (or center of the D0 gap), labeled $\Delta$E, is shown as an approximate indicator of the degree of non-magnetic Coulomb perturbation at the surface, which may not relate to the amount of deposited iron in the same way for samples with different bulk carrier densities. The partner-swapping connectivity observed in the simulation and data is a simple way by which new states can be added to the surface band structure without disrupting the surface conditions required by the Z$_2$ topological order of Bi$_2$Se$_3$ \cite{TIbasic}. The spin-splitting of topological surface bands is often discussed as a special case of the Rashba effect (e.g. Ref. \cite{TIbasic,FuHexagonal}), in which the ``Rashba" Hamiltonian term causes surface electronic states to be spin-split by an energy proportional to their momentum $\vec{k}$. Our data and numerical simulations show that this description is only accurate for Bi$_2$Se$_3$ in a small part of the Brillouin zone surrounding the Brillouin zone center, because at momenta further from the $\overline{\Gamma}$-point the electronic states pair off, and are nearly spin degenerate (see Fig. 3b). This can be understood because the origin of the topological insulator state in Bi$_2$Se$_3$ is a symmetry inversion that occurs at the $\Gamma$-point \cite{MatthewNatPhys,ZhangPred}, and the electronic states close to the Brillouin zone boundary are similar to those of topologically trivial materials.


Recent theoretical studies suggest that the physical environment of magnetic impurities on a topological surface is very different from the surface environment provided by a normal semiconductor such as silicon \cite{Biswas,zhangTImagImp,FerroSplitting}. When a non-magnetic crystal is doped with magnetic impurities, long-range magnetic ordering can come about as a result of itinerant electrons exchange-mediating the magnetic interaction. In a normal three dimensional material, electrons that interact with surface-deposited magnetic impurities are free to scatter away from the surface over a 2$\pi$ solid angle, and the two dimensional magnetic interactions therefore typically decay over several Angstroms (e.g. Ref. \cite{adatomExchange}). In topological insulators, surface state electrons are naturally confined to the surface in two dimensional Dirac cones, and it is suggested that interactions between deposited impurities can be mediated over many nanometers (see illustration in Fig. 4c) \cite{Biswas,zhangTImagImp}. The momentum-locked spin polarization of topological surface electrons causes them to support ferromagnetic order with an out-of-plane bias ($\vec{B}$ along the $\pm\hat{z}$ direction) at high impurity densities \cite{FerroSplitting,Biswas}, unlike direct dipole-dipole interactions which are much weaker and favor in-plane magnetic orientation.

Our numerical simulations have shown that the new D1 and D2 Dirac point electrons are localized deeper inside of the material than D0, and are expected to interact more weakly with electronic orbitals of the surface-deposited iron. Therefore, we focus on the D0 electrons to understand the effect of iron magnetism on the topological surface. A gap appears at the D0 Dirac point after heavy iron deposition on Samples $\#$1 and $\#$3, and the lower D0 Dirac cone acquires a buckled shape with a local energy minimum at zero momentum. The mass induced in the upper D0 Dirac cone after full deposition is approximately 0.1 M$_e$ (electron masses), and the surface band gap is similar to 100 meV. It is difficult to identify the exact point at which the band gap appears, likely because there is no long-range order of iron spins on the surface \cite{MerminWagner,UltrathinMag}, and differently ordered domains will yield differently gapped contributions to the photoemission signal.

Global ferromagnetism breaks time reversal symmetry, making it possible to induce a gap at the Dirac point, which is otherwise disallowed by the crystal symmetry. Fig. 4a shows the result of modifying a GGA numerical prediction of the D0 surface state by adding perturbative coupling to a surface layer of magnetic impurities with ferromagnetic out-of-plane order (equivalent to the Zeeman effect), yielding a dispersion that closely matches the experimental data. Details of this calculation and factors that can modify the surface state dispersion in numerical simulations will be discussed in an upcoming publication. We note that evidence of Landau levels is neither seen in our data nor expected from theory, as the D0 gap is due to hybridization with spin polarized Fe 3d orbitals rather than to coupling with the extremely weak magnetic field from iron dipole moments. If the D0 band gap were due to an isotropic out-of-plane magnetic field, a similar gap would be expected at the D1 and D2 Dirac points, which is clearly not observed. Zeeman-type magnetic symmetry breaking on topological Dirac surface electrons induces a band gap proportional to the out-of-plane magnetic moment, and in-plane moment shifts bands in momentum space which will cause the ARPES features to be broadened if the photon beam spot is much larger than the magnetic domain size \cite{FerroSplitting,FerroPosMassBiSe}. Based on the appearance of a gap at D0 in our data, we conclude that much of the iron-deposited crystal surface is occupied with out-of-plane ferromagnetically ordered domains.

The unusual effect observed in Fig. 2a, in which photoemission images become sharper as the D0 gap appears, is summarized in Fig. 4d in terms of the approximate band width of electronic states composing the D0 upper Dirac cone. Band width was measured through Lorentzian fitting at binding energies between the D0 Dirac point and the onset of the D1 band structure. Band width is inversely related to the electronic mean free path, and the reduction in band width as the gap appears is likely indicative of a magnetic disorder-to-order transition, leading to a reduction in scattering. In the special case of topological insulators, an alternative explanation for reduced band width might be that the magnetic moments of local ferromagnetic domains are initially oriented in arbitrary directions, and only gravitate strongly towards the z-axis at higher deposited densities for which an out-of-plane orientation may be more strongly energetically favored \cite{Biswas}. Reducing the prevalence of magnetic domains with in-plane moment will cause the surface state bands to appear narrower even if electronic mean free paths are unchanged, due to the effect of domain averaging. In either scenario, the close correspondence between reduced band width and gapping of the D0 Dirac point strongly indicates the occurrence of an out-of-plane magnetic ordering phase transition after approximately six minutes of iron deposition.

This apparent ordering transition driven by magnetic interactions that are mediated by the topological surface state is an indication of how strong topological Z$_2$ order changes the character of the material surface. The observations reported here, including out-of-plane surface magnetism and the appearance of new topological surface states, open a window into how topological surface states are formed and interact with perturbations. As such, they are significant for theoretical understanding of the formation of the topological insulator state in particular materials, and have direct implications for proposed devices utilizing Coulomb-charged or magnetic interfaces with topological insulators, such as capacitors \cite{ExcitCapacitor} and junctions \cite{palee,FerroSplitting,KaneDevice,ZhangDyon}.

\textbf{Methods summary:}
Angle resolved photoemission spectroscopy (ARPES) measurements were performed at the Advanced Light Source (LBNL, Berkeley) beamlines 12 and 10 with better than 15 meV energy resolution and overall angular resolution better than 1$\%$ of the Brillouin zone (BZ). Samples were cleaved and measured at 15$^o$K, in a vacuum maintained below 8$\times$10$^{-11}$ Torr. Momentum along the $\hat{z}$ axis is determined using an inner potential of 9.5 eV, consistent with previous photoemission investigations of undoped Bi$_2$Se$_3$ \cite{MatthewNatPhys, WrayCuBiSe}. Fe atoms were deposited using an e-beam heated evaporator at a rate of approximately 0.1$\AA$/minute. A quartz micro-balance supplied by Leybold-Inficon with sub-Angstrom sensitivity was used to calibrate the iron deposition flow rate. Surface and bulk state band calculations were performed for comparison with the experimental data, using the LAPW method implemented in the WIEN2K package (wien2k). Details of the theoretical calculation are similar to those in Ref. \cite{MatthewNatPhys, Matthew2008}.

\newpage

\begin{figure*}[t]
\includegraphics[width = 11cm]{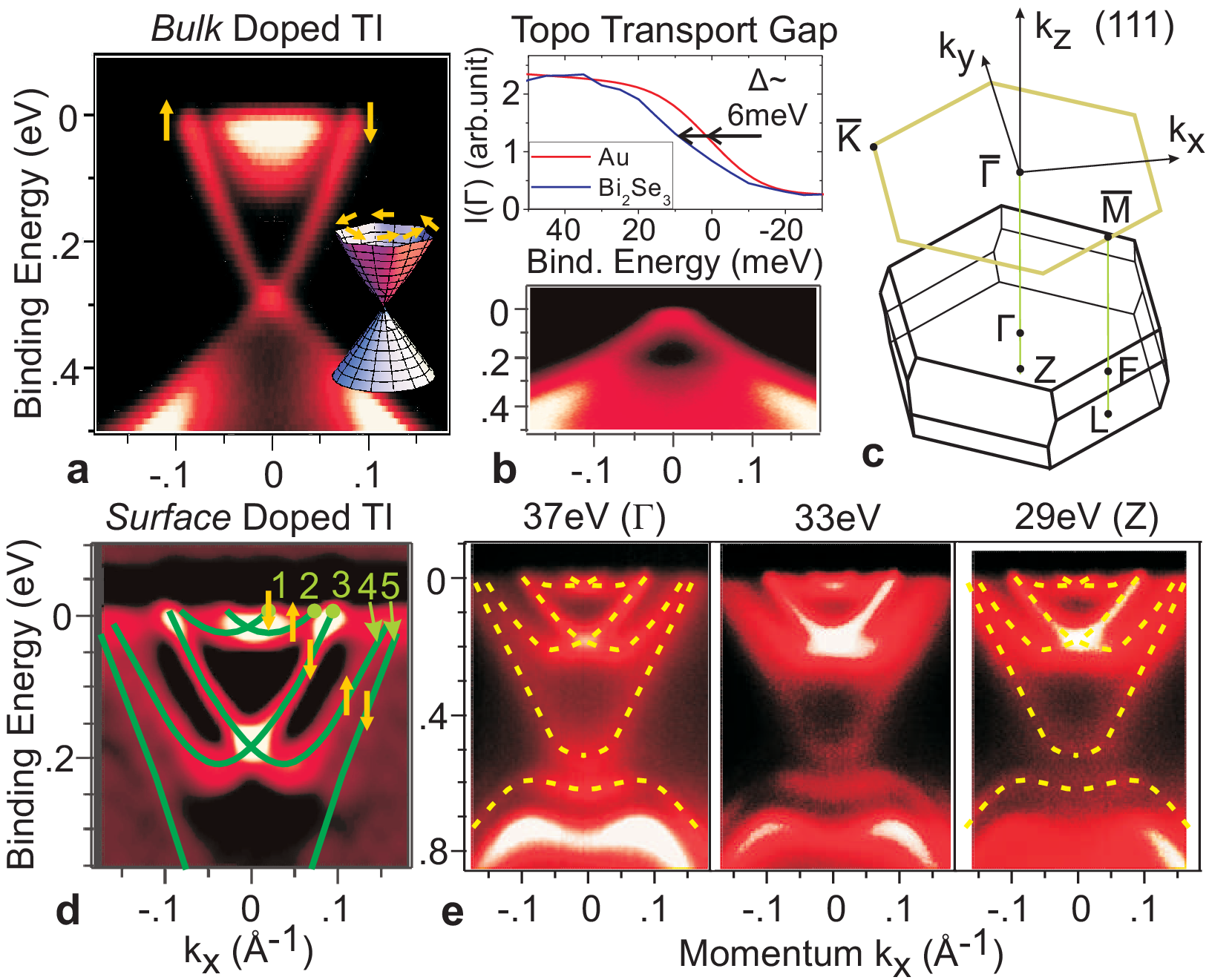}
\caption{{\bf{Iron deposition strongly modifies the topological surface}}:
\textbf{a}, Uniformly electron doped Bi$_2$Se$_3$ has a single surface state Dirac cone. \textbf{b} When the chemical potential of Bi$_2$Se$_3$ is lowered to the Dirac point by NO$_2$ deposition, a small gap is observed in the leading edge of ARPES intensity. This could be due to self-energy effects or magnetic impurities (Mn or Fe) induced effects. \textbf{c}, The hexagonal surface Brillouin zone of Bi$_2$Se$_3$ is drawn above a diagram of the three dimensional bulk Brillouin zone. \textbf{d}, A second derivative image of new surface states in Bi$_2$Se$_3$ (Sample $\#$1) after surface iron deposition is labeled with numerically predicted spin texture from Fig. 3b. \textbf{e}, Low energy features from \textbf{d} have no z-axis momentum dispersion. An early work on this was reported by Xia \textit{et.al.}, http://arxiv.org/abs/0812.2078 (2008).
}
\end{figure*}

\begin{figure*}[t]
\includegraphics[width = 9cm]{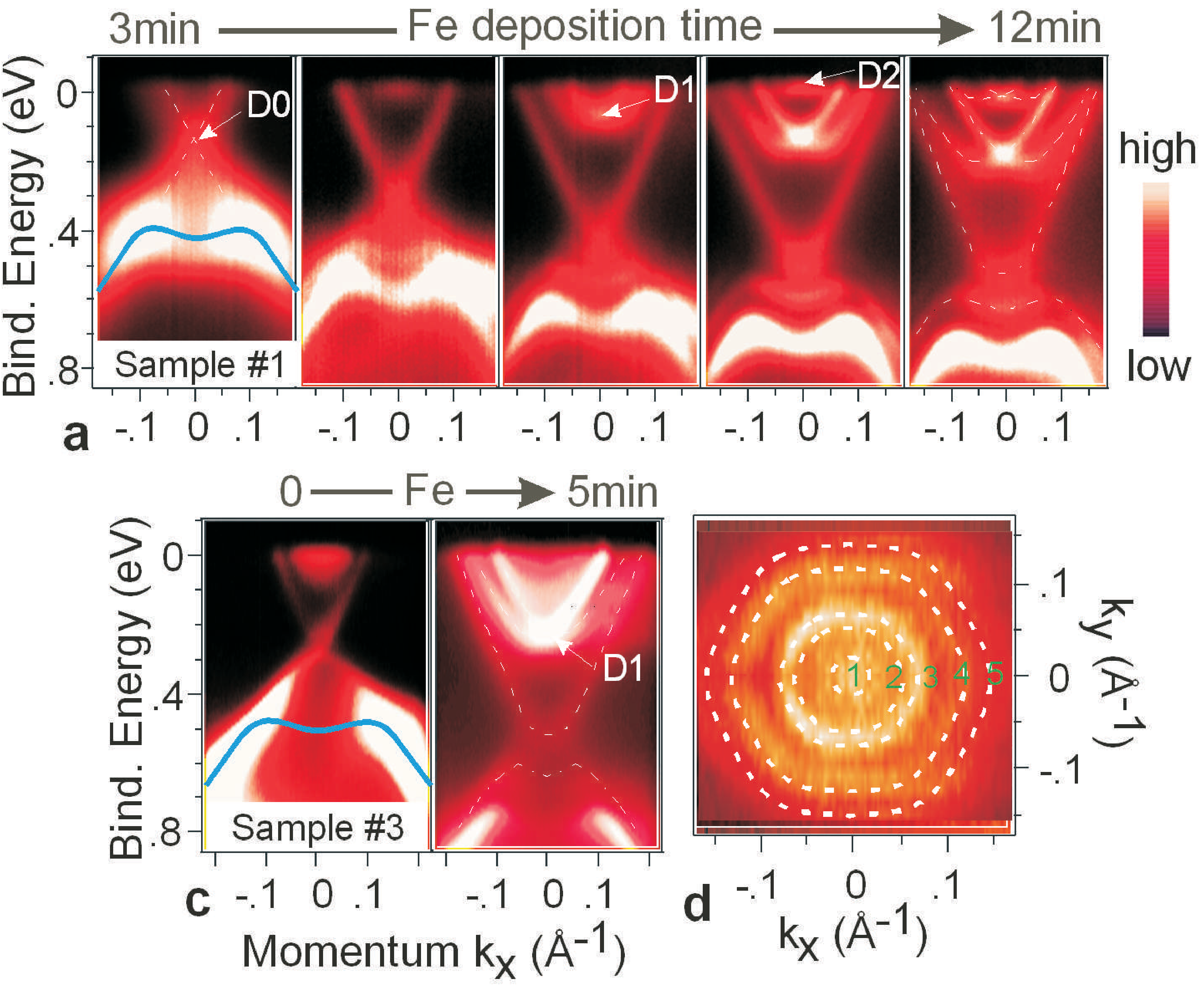}
\caption{{\bf{Iron doping creates five Dirac cones}}: \textbf{a}, Progressive surface doping of optimally insulating Ca$_x$Bi$_{2-x}$Se$_3$ (Sample $\#$1) causes the successive appearance of two new Dirac points (``D1" and ``D2"). \textbf{b}, Light surface doping of slightly n-type Bi$_2$Se$_{3.04}$ (Sample $\#$2) causes a new band to appear, without an obvious Dirac point. \textbf{c}, Iron deposition on as-grown n-type Bi$_2$Se$_3$ (Sample $\#$3) results in a similar final spectrum to panel \textbf{a}, but with only a single new Dirac point. \textbf{d}, A Fermi surface map of sample $\#$1 after heavy iron deposition shows successive concentric rings.}
\end{figure*}

\begin{figure*}[t]
\includegraphics[width = 17cm]{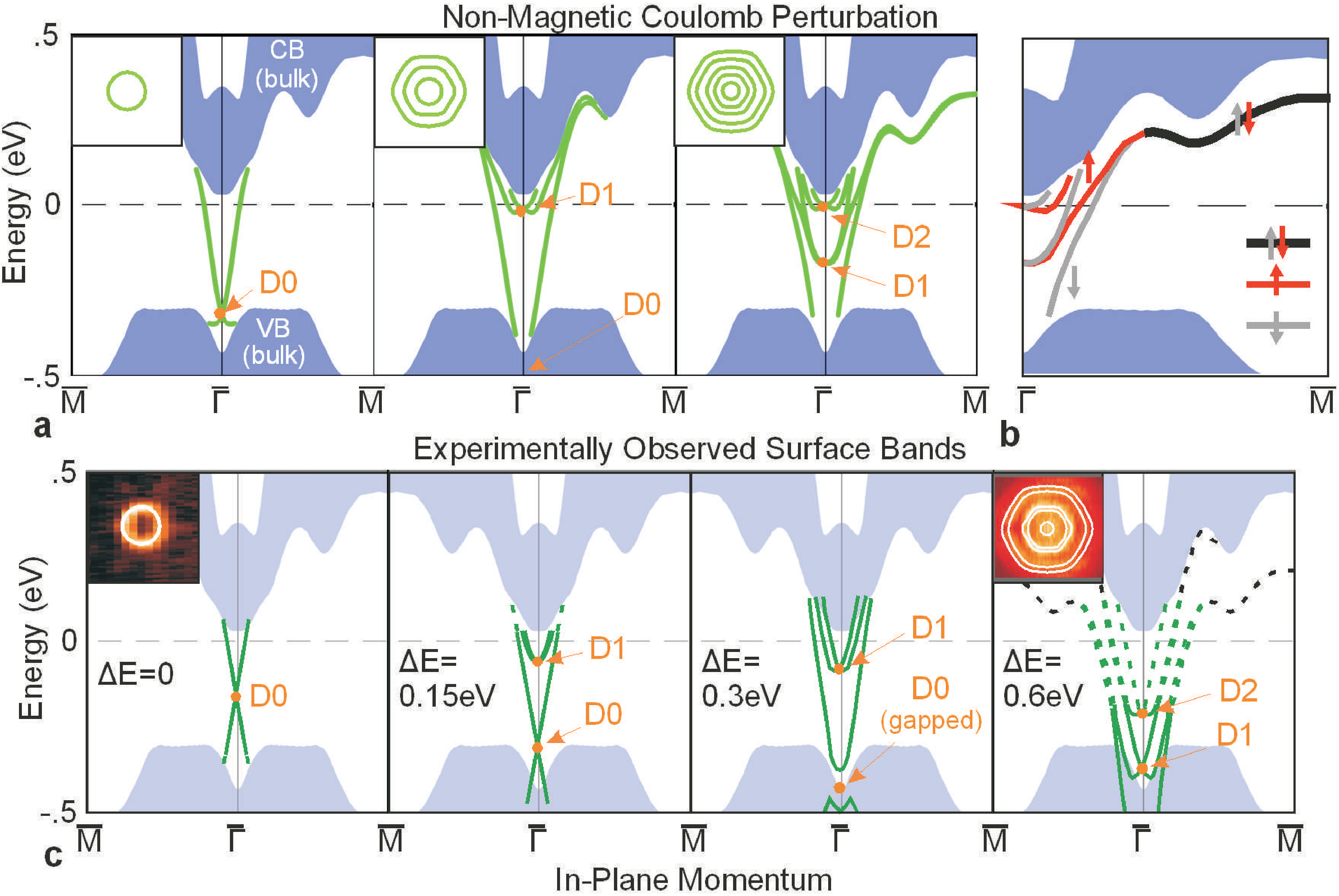}
\caption{{\bf{Magnetization dynamics of topological surface under Coulomb and magnetic perturbations}}: \textbf{a}, A numerical model of the Bi$_2$Se$_3$ surface state band structure shows the appearance of new Dirac points when the orbital energies are lowered by 0.68eV on (center) the first and (right) the first two quintuple layers of the lattice, simulating the non-magnetic Coulomb interaction with deposited Fe ions. Inset schematics show the Fermi surface evolution (not to scale). \textbf{b}, Spin texture for the rightmost panel of (a) is labeled with $\uparrow$ for right handed and $\downarrow$ for left handed chiral spin texture. \textbf{c}, Evolution of the surface band structure under doping is shown, based on white guides to the eye in Fig. 2. Panels are labeled with ($\Delta$E) the energy shift of the D0 Dirac point (or center of the D0 gap) relative to the D0 binding energy before surface deposition. Dashed lines show the expected dispersion of unoccupied bands, based on theory in panel (a), and an inset shows the Fermi surface map from Fig. 2d.}
\end{figure*}

\begin{figure*}[t]
\includegraphics[width = 9cm]{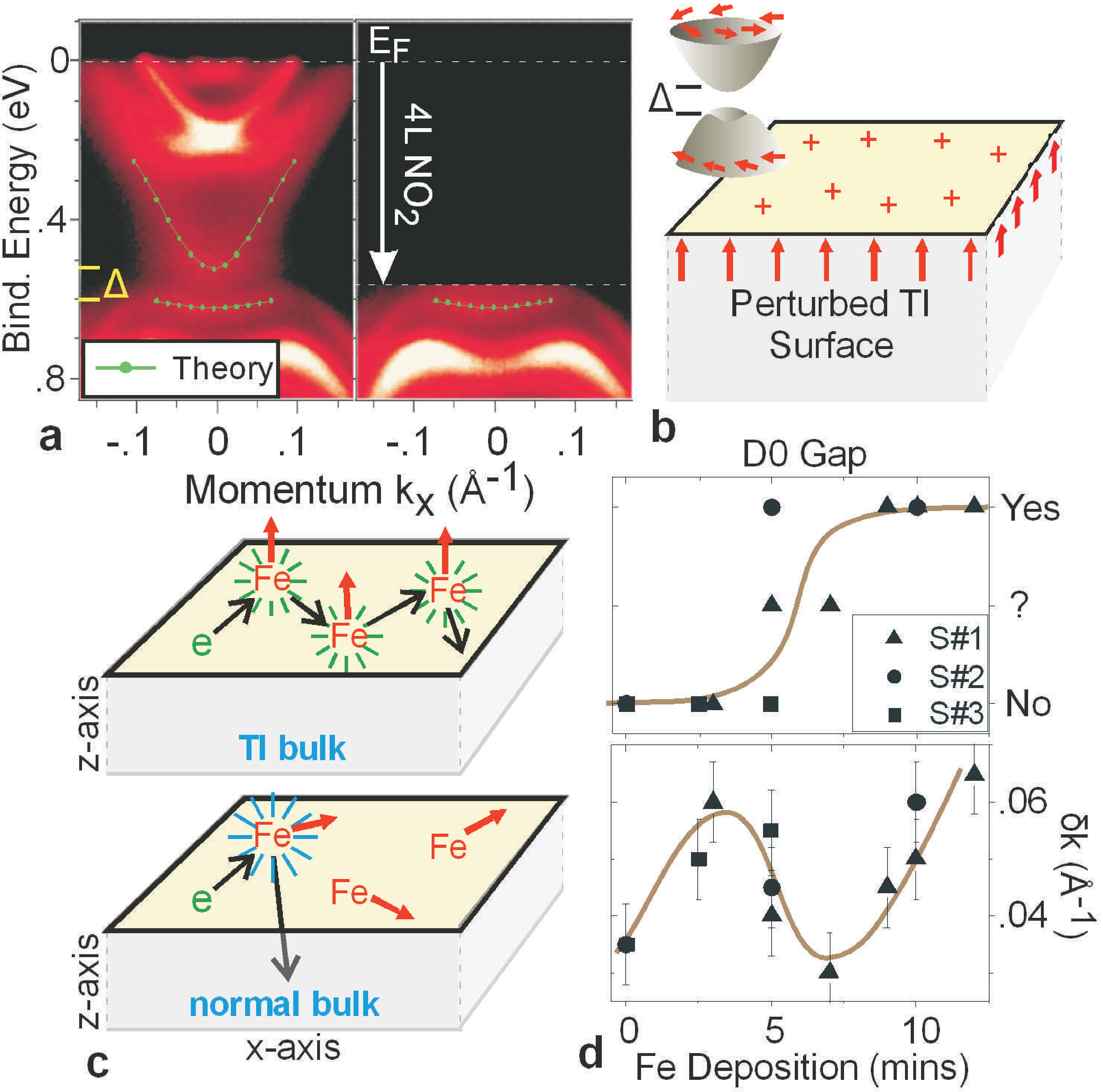}
\caption{{\bf{Observation of a surface phase transition under Coulomb perturbation and magnetism}}: \textbf{a}, (left) A z-axis magnetic perturbation causes (green) GGA predicted surface states to conform to the iron doped dispersion. (right) Fig-4a(right) is identical to Fig-4a(Left) showing where the Fermi level would be based on the experimental results in our earlier work in Ref-\cite{DavidTunable}. A surface insulator phase can be obtained by adding $\sim$4 Langmuirs \cite{DavidTunable} of non-magnetic NO$_2$ on the surface subsequent to Fe deposition. Figure shows a schematic (not data). \textbf{b}, A cartoon illustrates Fe deposition on the TI surface. \textbf{c}, A topological surface electron that scatters from surface magnetic impurities (iron) is likely to remain confined to the surface and interact with multiple Fe atoms in a short time, mediating strong magnetic interactions between the impurities. In normal semiconductors, electrons that interact with defects on the surface will quickly scatter back into the bulk. \textbf{d}, The development of a gap at the lowest energy (``D0") Dirac point is compared with (bottom) the half maximum peak width ($\delta$k) of connected states in the D0 upper Dirac cone. Error bars represent estimated uncertainty based on the results of different fitting techniques. An early work on this was reported by Xia \textit{et.al.}, http://arxiv.org/abs/0812.2078 (2008).}
\end{figure*}

\end{document}